\documentclass[jkps,twocolumn,fleqn,showpacs,showkeys]{revtex4}
\usepackage{graphicx}
\usepackage{amssymb}
\usepackage{amsmath}
\usepackage{bm}

\newcommand{\bea}{\begin{eqnarray}}
\newcommand{\eea}{\end{eqnarray}}
\newcommand{\be}{\begin{equation}}
\newcommand{\ee}{\end{equation}}
\newcommand{\no}{\nonumber}

\begin{document}
\setcounter{page}{0}
\title[]{Consistent Hamiltonian Reduction}
\author{Jong Hyuk \surname{Yoon}}
\email{yoonjh3404@gmail.com}
\affiliation{School of Physics,
Konkuk University, Seoul 143-701, Korea}
\date[]{}

\begin{abstract}

I show that the recently proposed (2+2) Hamiltonian reduction of Einstein's equations
of 4-dimensional spacetimes is consistent with general covariance.
The consistency proof is {\it extrinsic}, as it follows from the fact that
Hamilton's equations derived from the 
non-zero gravitational Hamiltonian are identical to the Ricci-flat condition
of 4-dimensional spacetimes in privileged coordinates.

 \end{abstract}



\maketitle

\renewcommand{\theequation}{\arabic{section}.\arabic{equation}}

\section{INTRODUCTION}{\label{a:intro}}
\setcounter{equation}{0}

Recently, I proposed the (2+2) Hamiltonian 
reduction\cite{ADM,adm60a,kuchar71,rovelli94,brown95,husain94,torre96,husain12}
of spacetimes of 4 dimensions under no symmetry assumptions
based on the (2+2) decomposition of the 4-dimensional spacetimes\cite{yoon92,yoon93a,yoon99a,yoon99c,yoon04}, 
where the area element of the cross-section of null hypersurfaces is treated as
the physical time, and the physical radius is defined by equipotential
surfaces on a given spacelike hypersurface of constant physical time.
In this Hamiltonian reduction, the Hamiltonian and momentum constraints are
simply the {\it defining} equations of the densities of the 
physical Hamiltonian and momentum of Einstein's theory 
in term of the conformal two-metric and its conjugate momentum\cite{kuchar71}.

As a consistency check of this Hamiltonian reduction, 
I will show that Hamilton's equations of motion that follow
from the non-zero gravitational Hamiltonian are identical to the Ricci-flat condition
\be
{\rm Ricci} =0                           \label{redshift}
\ee
of the 4-dimensional spacetime in privileged coordinates. 
This shows that the entire procedure of the Hamiltonian
reduction is consistent with Einstein's equations\cite{ADM,adm60a}.
The consistency proof should be regarded as an {\it extrinsic} one,  as the proof
relies on a comparison with Einstein's equations.
Thus, as a result of the Hamiltonian reduction, one is able to reformulate
Einstein's theory as a regular Hamiltonian system
of a $2\times 2$ symmetric uni-modular matrix and its trace-free
conjugate momentum.

This paper is organized as follows. In Section \ref{a:action} the (2+2) 
bundle formulation of Einstein's theory of 4-dimensional spacetimes is reviewed, 
and the concept of the covariant derivatives associated with the connections valued in the Lie algebra of the diffeomorphisms of the 2-dimensional vertical space is introduced. 
In Section \ref{a:hama} privileged spacetime coordinates are introduced 
by identifying certain functions of the gravitational phase space as
spacetime coordinates. 
In Section \ref{a:main}
all the equations obtained through the Hamiltonian reduction are summarized. 
In Section \ref{a:users} the Ricci-flat conditions in privileged coordinates 
are compared with Hamilton's equations of motion 
obtained from the non-zero gravitational Hamiltonian, and it is found 
that they are identical. 
In Section \ref{a:discussion} the nature of the privileged coordinates and their limitations are discussed.


\section{Review of the (2+2) formalism}{\label{a:action}}
\setcounter{equation}{0}

Let us recall that the metric in the (2+2) decomposition\cite{yoon92,yoon93a,yoon99a,yoon99c,yoon04,sachs62,dinverno78} of 
4-dimensional spacetimes can be written as 
\bea 
& & \hspace{-0.6cm}
ds^2 = 2dudv - 2hdu^2 +\tau\rho_{ab}
 \left( dy^a + A_{+}^{\ a} du + A_{-}^{\ a} dv \right)  \nonumber\\
& & \hspace{0.2cm} 
\times
\left( dy^b + A_{+}^{\ b} du + A_{-}^{\ b} dv  \right).          \label{yoon}
\eea
The horizontal vector fields $\hat{\partial}_{\pm}$ defined as
\be
\hat{\partial}_{\pm}:=\partial_{\pm} - A_{\pm}^{\ a} \partial_{a}
\ee
are orthogonal to the 2-dimensional spacelike surface
$N_{2}$ generated by $\partial_{a}$,
where
\be
\partial_{+}  = {\partial / \partial u},  \
\partial_{-}  = {\partial / \partial v},  \
\partial_{a}  = {\partial / \partial y^{a}}   \ (a=2,3).
\ee
The inner products $< \, , >$ 
of the basis vector fields $\{ \hat{\partial}_{\pm}, \partial_{a} \}$
are given by
\bea
& & \hspace{-0.4cm}
<\hat{\partial}_{+}, \ \hat{\partial}_{+}> = -2h, \hspace{0.2cm}
<\hat{\partial}_{+}, \ \hat{\partial}_{-}> = 1,  \hspace{0.2cm}
<\hat{\partial}_{-}, \ \hat{\partial}_{-}> =0, \nonumber\\
& &  \hspace{-0.4cm}
<\hat{\partial}_{\pm}, \ \partial_{a}> =0, \hspace{0.2cm}
<\partial_{a}, \ \partial_{b}> =\tau\rho_{ab}.
\eea
Thus, $\hat{\partial}_{-}$ is a null vector field because its norm is zero, 
whereas $\hat{\partial}_{+}$ can be either a spacelike, timelike, or null
vector field
depending on the sign of its norm $-2h$ 
that can be either positive, negative, or zero.
In this paper,  I assume $-2h>0$ so that 
$v= {\rm constant}$ is a spacelike hypersurface. The metric on $N_{2}$  is
$\tau\rho_{ab}$, where $\tau$ is the area element of $N_{2}$ and 
$\rho_{ab}$ is the conformal two-metric with $\det \rho_{ab}=1$.

As was shown in Ref. \cite{yoon04}, Einstein's equations  
can be obtained from the following action integral
\bea
& & \hspace{-0.4cm}
S=\int \!\! dv du d^{2} y \{ \pi_{\tau}\dot{\tau} + \pi_{h}\dot{h}
+ \pi_{a}\dot{A}_{+}^{\ a}    
+ \pi^{ab}\dot{\rho}_{ab}   \nonumber\\
& & 
 \hspace{0.5cm}
  -``1" \cdot C_{-} -``0" \cdot C_{+} - A_{-}^{\ a}C_{a} \},                              \label{bareaction}
\eea
where the overdot means $\partial_{-}$, and
$``1"$, $``0"$, and $A_{-}^{\ a}$ are Lagrange multipliers that enforce
the constraints  $C_{-}=0$, $C_{+}=0$, and $C_{a}=0$:
\begin{eqnarray}
& & \hspace{-1.2cm}  1. \
C_{-}:=  {1\over 2}\pi_{h}\pi_{\tau}
-{h\over 4\tau}\pi_{h}^{2}
-{1\over 2\tau}\pi_{h}D_{+}\tau
+{1\over 2\tau^{2}}\rho^{a b}\pi_{a}\pi_{b}\nonumber\\
& & \hspace{-0.5cm} 
-{\tau\over 8h}\rho^{a b} \rho^{c d}
(D_{+}\rho_{a c}) (D_{+}\rho_{b d})
-{1\over 2h \tau}
\rho_{a b}\rho_{c d}\pi^{a c}\pi^{b d}  \nonumber\\
& &  \hspace{-0.5cm} 
-{1\over 2h}\pi^{a c}D_{+}\rho_{a c}
-\tau {\rm R}_{(2)}  + D_{+}\pi_{h}
- \partial_{a}(\tau^{-1}\rho^{a b} \pi_{b}) \nonumber\\
& &  
=0,  \label{C}
\end{eqnarray}
\begin{eqnarray}
& & \hspace{-1.2cm} 2. \ 
C_{+}:=\pi_{\tau}D_{+}\tau + \pi_{h}D_{+}h + \pi^{a b}D_{+}\rho_{a b} \nonumber\\
& &  \hspace{-0.5cm} 
- 2D_{+}(
h \pi_{h}  +  D_{+} \tau )
+2\partial_{a}(h \tau^{-1}\rho^{a b}\pi_{b}
 +\rho^{a b}\partial_{b}h )\nonumber\\
& & 
=0,         \label{C+}
\end{eqnarray}
\begin{eqnarray}
& & \hspace{-1.2cm} 3. \
C_{a}:=\pi_{\tau}\partial_{a}\tau +\pi_{h}\partial_{a}h
+\pi^{b c}\partial_{a}\rho_{b c}
-2\partial_{b}( \rho_{ac}\pi^{bc})  \nonumber\\
& &  \hspace{-0.5cm} 
-D_{+}\pi_{a} - \partial_{a}(\tau \pi_{\tau}) =0.   \label{CA}
\end{eqnarray}
Here, $ {\rm R}_{(2)}$ is the Ricci scalar of $N_{2}$, 
and the diff$N_{2}$-covariant derivative\cite{yoon04} of a tensor density
$q_{a b }$ with weight $w$ is defined as
\bea
& & \hspace{-0.4cm}
D_{\pm}q_{a b}:= \partial_{\pm}q_{a b}
-[A_{\pm}, \ q]_{{\rm L}a b}  
= \partial_{\pm}q_{a b} - A_{\pm}^{\ c}\partial_{c}q_{ab} \nonumber\\
& &
-q_{cb}\partial_{a}A_{\pm}^{\ c}
-q_{ac}\partial_{b}A_{\pm}^{\ c}    
-w (\partial_{c}A_{\pm}^{\ c})q_{ab},
            \label{covdiff}
\eea
where $[A_{\pm}, \ q]_{{\rm L} a b}$ is the Lie
derivative of $q_{ab}$ along $A_{\pm}:=A_{\pm}^{\ a}\partial_{a}$.
For instance, the diff$N_{2}$-covariant derivatives of
$\tau$, $\rho_{a b}$, and $h$, which are tensor densities with 
weight $1$, $-1$, and $0$ with respect to the diff$N_{2}$ transformations, respectively,
are given by
\bea 
& & \hspace{-1cm}
D_{\pm}\tau=\partial_{\pm}\tau
-A_{\pm}^{\ a}\partial_{a}\tau
-(\partial_{a} A_{\pm}^{\ a}) \tau,      \label{walleye}\\
& &  \hspace{-1cm}
D_{\pm}\rho_{a b}=\partial_{\pm} \rho_{a b}
-A_{\pm}^{\ c}\partial_{c} \rho_{a b}
-\rho_{c b}\partial_{a}A_{\pm}^{\ c}
-\rho_{a c}\partial_{b}A_{\pm}^{\ c} \no\\
& & \hspace{0.3cm}
+ (\partial_{c}A_{\pm}^{\ c})\rho_{a b},   \label{bass}\\
& & \hspace{-1cm}
D_{\pm}h= \partial_{\pm}h - A_{\pm}^{\ a}\partial_{a}h,   \label{cut}        
\eea
and the diff$N_{2}$-covariant field strength $F_{+-}^{\ \ a}$ is defined as 
\bea
& &  \hspace{-1cm}
F_{+-}^{\ \ a}:=\partial_{+} A_{-} ^ { \ a}-\partial_{-} A_{+} ^ { \ a} 
- [A_{+}, \ A_{-}]^{a}   \no\\
& & \hspace{-0.1cm}
=\partial_{+} A_{-} ^ { \ a}-\partial_{-} A_{+} ^ { \ a} 
-A_{+}^{\ b}\partial_{b}  A_{-} ^ { \ a} 
+A_{-}^{\ b}\partial_{b}  A_{+} ^ { \ a}.  \label{throat}
\eea

The canonical momenta $\pi_{I}=(\pi_{\tau}, \pi_{h},\pi_{a}, \pi^{ a b})$ conjugate to the configuration variables
$q^{I}=(\tau, h, A_{+}^{\ a}, \rho_{ab})$ are defined  as follows,
\bea 
 & & \hspace{-0.5cm}
\pi_{h}:=2D_{-}\tau,   \label{broker}\\
& &  \hspace{-0.5cm}
\pi_{\tau} :=2h \tau^{-1}D_{-}\tau +2D_{-}h  + \tau^{-1}D_{+}\tau , \label{jack}\\
& &  \hspace{-0.5cm}
\pi_{a}:=-\tau^{2} \rho_{a b}F_{+-}^{\ \ b}, \label{heart}\\
& &  \hspace{-0.5cm}
\pi^{a b}:=-h \tau \rho^{a c}\rho^{b d}D_{-}\rho_{c d}
-{1\over 2}\tau \rho^{a c}\rho^{b d}D_{+}\rho_{c d}.      \label{after}
\eea
The diff$N_{2}$-covariant derivatives of the conjugate momenta $\pi_{\tau}$, 
$\pi_{h}$, $\pi_{a}$, and $\pi^{ab}$, which are tensor densities of weights
$0, 1, 1, 2$, respectively, are given by
\bea 
& &  \hspace{-1.2cm}
D_{\pm} \pi_{\tau}=\partial_{\pm} \pi_{\tau}  
-  A_{\pm}^{\ a}\partial_{a}\pi_{\tau},                         \label{close}\\
& & \hspace{-1.2cm}
 D_{\pm} \pi_{h}=\partial_{\pm} \pi_{h}  
-  A_{\pm}^{\ c}\partial_{c}\pi_{h} -(\partial_{c} A_{\pm}^{\ c}) \pi_{h}, \label{hill}\\
& & \hspace{-1.2cm}
D_{\pm} \pi_{a}=\partial_{\pm} \pi_{a}  
-  A_{\pm}^{\ c}\partial_{c}\pi_{a}  -\pi_{c}\partial_{a}A_{\pm}^{\ c}
- (\partial_{c} A_{\pm}^{\ c}) \pi_{a},       \label{bank}\\
& & \hspace{-1.2cm}
D_{\pm} \pi^{ab}=\partial_{\pm}\pi^{ab}
-  A_{\pm}^{\ c}\partial_{c}\pi^{ab}   +\pi^{cb}\partial_{c}A_{\pm}^{\ a}
+ \pi^{ac}\partial_{c}A_{\pm}^{\ b}    \no\\
& & \hspace{0cm}
-2 (\partial_{c} A_{\pm}^{\ c}) \pi^{ab}.     \label{dep}
\eea
Because the conformal two-metric $\rho_{ab}$ has a unit determinant, 
its derivatives satisfy the equations
\begin{equation}
\rho^{bc}\partial_{\pm}\rho_{bc}=\rho^{bc}\partial_{a}\rho_{bc}
=\rho^{bc}D_{\pm}\rho_{bc}=0.  \label{module}
\end{equation}
From these equations and Eq. (\ref{after}), the conjugate momentum $\pi^{a b}$ 
is traceless:
\begin{equation}
\rho_{ab}\pi^{a b}=0.           \label{side}
\end{equation}

\section{Hamiltonian reduction} {\label{a:hama}}
\setcounter{equation}{0}

Let us define a {\it potential} function $R$ and its conjugate momentum
$\pi_{R}$ as\cite{kuchar71}
\bea
& & 
\partial_{+}R:=-h\pi_{h},  \label{iden}\\
& & 
\pi_{R}=-\partial_{+}{\ln}(-h),      \label{pir}
\eea
respectively. The transformation from  $(h, \pi_{h})$ to 
$(R, \pi_{R})$ is clearly a {\it canonical} transformation, as it changes
the action integral by total derivatives only.
If we impose the constraints $C_{+}=0$ and $C_{a}=0$  and choose the 
Lagrange multiplier $A_{-}^{\ a}=0$, then
the action in Eq. (\ref{bareaction}) becomes
\bea 
& & \hspace{-0.5cm}
S=\int \!\! dv du d^{2} y \{ \pi_{\tau}\dot\tau + \pi_{R}\dot R
+ \pi_{a}\dot A_{+}^{\ a} + \pi^{ab}\dot\rho_{ab} -C_{-} \}           \nonumber\\
& & \ + \ {\rm total \ derivatives},  \label{action1}
\eea
where the constraint $C_{-}$ in the new variables is given by
\bea 
& &  \hspace{-1.2cm}
C_{-}=-{ 1\over 2h}\pi_{\tau} \partial_{+}R
- {1\over 4h \tau} (\partial_{+}R)^{2}  
+{1\over 2h\tau}(D_{+}\tau)(\partial_{+}R) \nonumber\\
& & \hspace{-0.5cm}
-{1\over h}D_{+} (\partial_{+}R) + {1\over h}(\partial_{+}R) \partial_{+} \ln (-h) \nonumber\\
& & \hspace{-0.5cm}
-{1\over h}(\partial_{+}R )A_{+}^{\ a} \partial_{a} \ln (-h)  \nonumber\\
& & \hspace{-0.5cm}
 -\tau {\rm R}_{(2)} 
+{1\over 2\tau^{2}}\rho^{ab}\pi_{a}\pi_{b}   -\partial_{a}( \tau^{-1}\rho^{ab} \pi_{b} )  \nonumber\\
& & \hspace{-0.5cm}
-{1\over 2h \tau}
\rho_{a b}\rho_{c d}\pi^{a c}\pi^{b d}
-{\tau\over 8h}\rho^{a b} \rho^{c d}
(D_{+}\rho_{a c}) (D_{+}\rho_{b d}) \nonumber\\
& & \hspace{-0.5cm}
-{1\over 2h}\pi^{a c}D_{+}\rho_{a c}  =0.   \label{newham}
\eea 
The function $h$ still appears in Eq. (\ref{newham}), but because
$h$ is related to $R$ and $\pi_{R}$ by Eqs. (\ref{iden}) and (\ref{pir}), 
the constraint $C_{-}$ given by  Eq. (\ref{newham}) may be viewed as a function of new variables
$(\tau, Q^{I}; \pi_{\tau},  \Pi_{I})$, where 
$Q^{I}=(R, A_{+}^{\ a}, \rho_{ab})$ and  $\Pi_{I}=(\pi_{R}, \pi_{a}, \pi^{ab})$.
Notice that the first term in  Eq. (\ref{newham}) is linear in $\pi_{\tau}$, and all the remaining terms are independent of  $\pi_{\tau}$. 
Thus, the equation of motion for $\tau$ is given by
\be
\dot{\tau}=\int \!\! du d^{2}y  {\delta C_{-} \over \delta \pi_{\tau}}
=-{1\over 2h}\partial_{+}R. \label{taueom}
\ee
Now, recall that $\tau=\tau(v,u,y^{a})$. 
If this equation is solved for $v$, then $v$ becomes a function of $(\tau,u,y^{a})$ and consequently, $Q^{I}=( R$, $A_{+}^{\ a}$, $\rho_{ab} )$ may be regarded 
as functions of $(\tau,u,y^{a})$,
\be
Q^{I}=Q^{I}(\tau,u,y^{a}).
\ee
Therefore, it follows that
\bea
& & \hspace{-1cm}
\dot{Q}^{I}={\partial \tau \over \partial v} {\partial {Q}^{I}\over \partial \tau} 
+ {\partial u \over \partial v} {\partial {Q}^{I}\over \partial u} 
+{\partial y^{a} \over \partial v} {\partial {Q}^{I} \over \partial y^{a}}  \nonumber\\
& &\hspace{-0.45cm}
=\dot{\tau}\partial_{\tau}{Q}^{I}                    \label{cookie}
\eea
because ${\partial u /\partial v}={\partial y^{a} /\partial v}=0$. If Eq. (\ref{taueom}) is used, then $C_{-}$ can be written as
\be
C_{-}=-({2h \over \partial_{+}R})\dot{\tau} C_{-}.    \label{muse}
\ee
By Eqs. (\ref{cookie}) and (\ref{muse}), 
the action in Eq. (\ref{action1}) becomes
\bea
& &  \hspace{-1.2cm}
S=\int \!\! dvdu d^{2} y \dot{\tau} \{ 
\pi_{\tau}  +\pi_{R}\partial_{\tau} R +  \pi_{a}\partial_{\tau} A_{+}^{\ a}  \nonumber\\
& & 
+ \pi^{ab}\partial_{\tau}\rho_{ab} +({2h \over \partial_{+}R})C_{-} \}  \nonumber\\
& &  \hspace{-1.2cm} 
= \! \! \int \!\! d\tau du d^{2} y \{
\pi_{R}\partial_{\tau} R +  \pi_{a}\partial_{\tau} A_{+}^{\ a}
+ \pi^{ab}\partial_{\tau}\rho_{ab}
- C_{(1)}   \}   \nonumber\\
& &   \hspace{-1.2cm}
=\int \!\! d\tau du d^{2} y \{
\sum_{I} \Pi_{I}\partial_{\tau}{Q}^{I} -C_{(1)} \},  \label{action2}
\eea
where the substitution 
\be
dv \dot{\tau} \longrightarrow d\tau
\ee
are made in the second line, and $C_{(1)}$ is defined as
\be
C_{(1)}=- ({2h\over \partial_{+}R})C_{-} - \pi_{\tau}.         \label{box}
\ee
Notice that if the constraint $C_{-} =0$ is imposed,  the function $C_{(1)}$ reduces to
\be
C_{(1)}= -\pi_{\tau},     \label{plastic}
\ee
which is the non-zero Hamiltonian density associated with 
the physical time $\tau$.

The second step in the Hamiltonian reduction consists of identifying
arbitrarily specifiable coordinate $u$  as
\be
u=R,   \label{ur} 
\ee
and choosing arbitrary labels $y^{a}$ of $N_{2}$ as
\be
y^{a}=Y^{a}    \label{yy}
\ee
such that  the ``shift" vector $A_{+}^{\ a}$ is zero,
\be
A_{+}^{\ a}=0.                         \label{zerow}
\ee
For the class of spacetimes whose
spatial topology is either $N_{2}\times \mathbb{R}$ or $N_{2}\times S^{1}$, where
$N_{2}$ is a compact 2-dimensional space with the genus $g$, 
the condition in Eq. (\ref{zerow}) can always be satisfied by relabeling 
$N_{2}$ such that $Y^{a}={\rm constant}$ is
normal to $R={\rm constant}$ at each point on $\Sigma_{\tau}$.
If we continue to label $N_{2}$ in this way, then the ``shift"  $A_{+}^{\ a}$ 
can always be made zero for the class of spacetimes 
under consideration (see FIG. 1).
Thus, we will work in the privileged coordinates $X^{A}:=(\tau, R, Y^{a})$,
which satisfy the coordinate conditions\cite{kuchar71}
\be
{\partial X^{A} \over \partial X^{B}}= \delta_{B}^{\, A}.   \label{nate}
\ee
Then, it follows from Eq. (\ref{taueom}) that
\be
\dot{\tau}=-{1\over 2h} > 0,
\ee
which means that $\tau$ is a monotonically increasing function of 
the affine parameter $v$ of the out-going null vector field. 
\begin{figure}
\vspace{-1.8in}
\hspace{-2in}
\begin{center}
\includegraphics[width=5in]{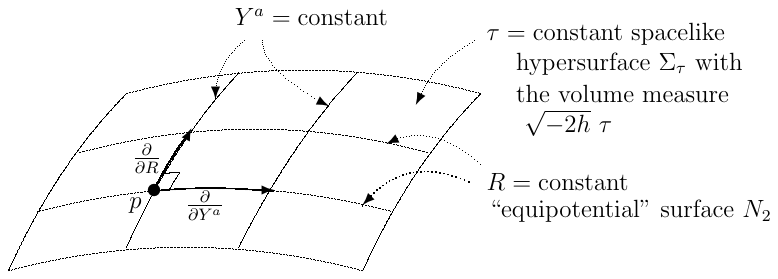}
\vspace{-4.4in}
\caption{\label{figure1} On the $R={\rm constant}$ ``equipotential" surface $N_{2}$
on $\Sigma_{\tau}$,  the coordinates $Y^{a}$ are introduced such that 
$Y^{a}={\rm constant}$ is
normal to $N_{2}$ at each point $p$ on $\Sigma_{\tau}$. Then, the ``shift" vector $A_{+}^{\ a}$ can always be made zero at $p \in \Sigma_{\tau}$.}
\end{center}
\vspace{-0.5cm}
\end{figure}
Hamilton's equations of motion follow from the variational principle
of the action integral (\ref{action2}): 
\bea
& & \partial_{\tau}Q^{I}=\int_{\Sigma_{\tau}} \!\!\!\! d u d^{2}y
{\delta C_{(1)} \over \delta \Pi_{I}}|_{u=R, y^{a}=Y^{a} },  \\
& & 
\partial_{\tau}\Pi_{I}=-\int_{\Sigma_{\tau}} \!\!\!\! d u d^{2}y
{\delta C_{(1)} \over \delta Q^{I}}|_{u=R, y^{a}=Y^{a}  },
\eea
with $Q^{I}=(R, A_{+}^{\ a},  \rho_{ab})$ and $\Pi_{I}=(\pi_{R}, \pi_{a},  \pi^{ab})$, and $\Sigma_{\tau}$ is a spacelike hypersurface defined by $\tau={\rm constant}$. 


\section{Hamilton's equations of motion}{\label{a:main}}
\setcounter{equation}{0}

In the following I present Hamilton's equations of motion in privileged 
coordinates:\\
{\it Einstein's evolution equations}
\bea
& & \hspace{-0.8cm}
1. \  {\partial R \over \partial \tau}=0 \Rightarrow  \nonumber\\
& &  \hspace{-0.4cm}
\tau  {\rm R}_{(2)}={1\over 2}\tau^{-2}\rho^{ab}\pi_{a}
\pi_{b}-\partial_{a}(\tau^{-1}\rho^{ab}\pi_{b})   \label{topo} \\
& & \hspace{-0.8cm}
2. \ {\partial A_{+}^{\ a} \over \partial \tau}=0 \Rightarrow  \nonumber\\
& &   \hspace{-0.4cm}
\tau^{-1}\pi_{a}=-\partial_{a}{\ln}(-h)  \label{parwa}  \\
& & \hspace{-0.8cm}
3. \ {\partial \tau\over \partial \tau}=1\\
& & \hspace{-0.8cm}
4.  \ {\partial \ {\ln}(-h) \over \partial \tau}=H - {1\over \tau}\\
& & \hspace{-0.8cm}
5.  \ {\partial \pi_{a}\over \partial \tau}=2\tau^{-1}\pi_{a} +
(\pi^{bc}+ {\tau\over 2} 
\rho^{bd}\rho^{ce}\partial_{R}\rho_{de})\partial_{a}\rho_{bc}\nonumber\\
& & \hspace{-0.4cm}
-\partial_{b}(2\pi^{bc}\rho_{ac}+\tau \rho^{bc}\partial_{R}\rho_{ac}) \label{parpi}\\
& &   \hspace{-0.8cm}
6.  \  {\partial \pi_{\tau}\over \partial \tau}= {1\over 2}\tau^{-2} 
+ \tau^{-2}\rho_{ab}\rho_{cd}\pi^{ac}\pi^{bd} \nonumber\\
& & \hspace{-0.4cm}
-{1\over 4}\rho^{ab}\rho^{cd}(\partial_{R}\rho_{ac})(\partial_{R}\rho_{bd}) 
-2\tau^{-2}\partial_{a}(h\rho^{ab}\pi_{b})  
\label{volvopitau}\\
& & \hspace{-0.8cm}
7.\   
\ {\partial\over \partial \tau}  \rho_{ab} =2 \tau^{-1}\rho_{a c}\rho_{b d}\pi^{cd}
+\partial_{R}\rho_{ab}                  \label{veryeasy}\\ 
& & \hspace{-0.8cm}
8.\  
\ {\partial\over \partial \tau} \pi^{ab}
=-2\tau^{-1}\rho_{c d}\pi^{a c}\pi^{b d}  + \partial_{R}\pi^{ab}   \no\\
& &  \hspace{ -0.5cm}
-{\tau\over 2}\rho^{a i}\rho^{b j}\rho^{c k} 
(\partial_{R}\rho_{i c})(\partial_{R}\rho_{j k})  
+{\tau\over 2}\rho^{ac}\rho^{b  d} (\partial_{R}^{2}\rho_{c d}) \no\\
& &  \hspace{ -0.7cm}
+2h  \rho^{a c}\rho^{b d} \{ 
{\rm R}_{(2)c d} - {1\over 2}\tau^{-2}\pi_{c}\pi_{d} 
+ \nabla_{c}^{(2)}(\tau^{-1} \pi_{d})  \}.         \label{getzen}
\eea
{\it Einstein's constraint equations}
\bea
& &  \hspace{-1cm}
9.\ C_{-}=0  \Rightarrow 
\pi_{\tau} = -H +2\partial_{R} \ln (-h)      \label{pitau}\\
& &\hspace{-0.6cm}
({\rm definition \ of \ physical \ Hamiltonian})        \nonumber
\eea
\bea
& & \hspace{-1cm}
10.\ C_{+}=0 \Rightarrow \pi_{R}=-\pi^{ab}\partial_{R}\rho_{ab}  \label{radi} \\
& &  \hspace{-0.6cm}
({\rm definition  \ of \ physical \ linear \ momentum})    \nonumber
\eea
\bea
& & \hspace{-1cm}
11.\ C_{a}=0  \Rightarrow \tau^{-1}{\pi}_{a}
= - \pi^{bc}\partial_{a}\rho_{bc}  +2\partial_{b}(\pi^{bc}\rho_{ac})   \nonumber\\
& & \hspace{2.6cm}
-\tau\partial_{a}( H+ \pi_{R})   \label{angko}\\
& &  \hspace{-0.6cm}
({\rm definition \ of \ physical \ angular \ momentum})   \nonumber 
\eea
{\it Superpotential} \  ${\ln}(-h)$
\bea
& &  \hspace{-1cm}
12.\ \partial_{\tau}  {\ln}(-h) =H - \tau^{-1}    \label{hstar}\\
& &  \hspace{-1cm}
13.\  -\partial_{R}  {\ln}(-h) =\pi_{R}    \label{pira}\\
& & \hspace{-1cm}
14.\  -\partial_{a}  {\ln}(-h) = \tau^{-1}\pi_{a}   \label{piwa}
\eea
{\it Integrability conditions}
\bea
& & \hspace{-1cm}
15.\ \partial_{R}(\tau^{-1}\pi_{a})=\partial_{a}\pi_{R}    \label{drtau}\\
& & \hspace{-1cm}
16.\ \partial_{\tau}\pi_{R} = -\partial_{R}H  \label{partaur}\\
& & \hspace{-1cm}
17.\  \partial_{\tau} (\tau^{-1}\pi_{a})=-\partial_{a}H \label{partautau}
\eea
In the above equations, $H$ is defined as
\bea
& & \hspace{-1cm}
18.\  H =   {1\over 2\tau} 
+{1\over \tau} \rho_{a b}\rho_{c d}\pi^{a c}\pi^{b d}  \nonumber\\
& &  \hspace{0cm}
+{1\over 4}\tau \rho^{a b} \rho^{c d}
(\partial_{R}\rho_{a c}) (\partial_{R}\rho_{b d})  
+\pi^{a c}\partial_{R}\rho_{a c}       
 \nonumber\\
& &  \hspace{0cm}
\geq  {1\over 2 \tau} .   \label{trueham}     
\eea
The spacetime metric in these privileged coordinates becomes
\be \hspace{-0.6cm}
19.\ ds^2 = -4h dR d\tau - 2h dR^2 +\tau\rho_{ab} dY^{a}  dY^{b}.
              \label{yoonspacetime}
\ee


\section{Consistency proof}{\label{a:users}}
\setcounter{equation}{0}

\subsection{Einstein's Equations in Privileged Coordinates}{\label{a:ein}}
Now, Einstein's equations 
\be
{\rm Ricci} =0
\ee
are computed in privileged coordinates $X^{A}=(\tau, R, Y^{a})$, and the equations summarized in Section \ref{a:main} will be shown to be
the first-order form of Einstein's equations.
Let the spacetime metric in Eq. (\ref{yoonspacetime}) be written as
\bea
& & 
ds^2 = 2e^{\phi}d\tau  dR  +e^{\phi} dR^2 
+\tau\rho_{ab} dY^{a}  dY^{b}        \nonumber\\
&  & \hspace{0.6cm}
=g_{AB}dX^{A}dX^{B},              \label{las}
\eea
where $e^{\phi} $ is defined as
\be
e^{\phi}:= -2h >0.             \label{respect}
\ee
In the metric in Eq. (\ref{las}), the fields $\phi$ and $\rho_{ab}$ are functions
of all the coordinates $(\tau, R, Y^{a})$, as no isometries are assumed.

Let $X^{A}=(X^{\mu}, Y^{a}) $, where $X^{\mu}=( X^{0}, X^{1}) :=(\tau, R)$. 
In the basis 
\be
\partial_{A}=
\big( {\partial \over \partial X^{\mu}}, {\partial \over \partial Y^{a}} \big), 
\ee
the metric can be written as
\be
g_{AB}=
\begin{pmatrix}
\gamma_{\mu\nu}   & 0  \\
0  &  \gamma_{ab} 
\end{pmatrix}
=\begin{pmatrix}
e^{\phi} S_{\mu\nu}   & 0  \\
0  &  \tau\rho_{ab} 
\end{pmatrix},  
\ee
where $S_{\mu\nu}$ is  defined as
\be
S_{\mu\nu}=
\begin{pmatrix}
0  &   1\\
1  &   1
\end{pmatrix}.
\ee
The inverse metric is given by
\be
g^{AB}=
\begin{pmatrix}
\gamma^{\mu\nu}   & 0  \\
0  &  \gamma^{ab} 
\end{pmatrix}
=
\begin{pmatrix}
e^{-\phi} S^{\mu\nu}   & 0  \\
0  &  \tau^{-1}\rho^{ab} 
\end{pmatrix},
\ee
where $S^{\mu\nu}$ is the inverse of  $S_{\mu\nu}$, 
\be
S^{\mu\nu}=
\begin{pmatrix}
-1 &   1\\
1  &   0
\end{pmatrix},
\ee
which satisfies the condition
\be
S^{\mu\nu}S_{\nu\alpha}=\delta_{\alpha}^{\, \mu}.
\ee
The Levi-Civita connections $\Gamma_{AB}^{\ \ \,C}$ and the Ricci tensor 
${\rm R}_{AC}$ are given by
\bea  
& & \hspace{-1.3cm}
\Gamma_{AB}^{\ \ \,C}={1\over 2}g^{CD} \big( 
\partial_{A}g_{BD} +\partial_{B}g_{AD}  - \partial_{D}g_{AB}  \big),  \label{emma}\\
& &  \hspace{-1.3cm}
{\rm R}_{AC}=\partial_{A}\Gamma_{BC}^{\ \ \, B} \!
-\partial_{B}\Gamma_{AC}^{\ \ \, B}  \!
+\Gamma_{AE}^{\ \ \, B} \Gamma_{BC}^{\ \ \, E} \!
-\Gamma_{BE}^{\ \ \, B} \Gamma_{AC}^{\ \ \, E},   \label{nuell}
\eea
respectively. Explicitly, the connection coefficients are found as follows:
\bea 
& &  \hspace{-1cm} 
(1) \ 
\Gamma_{\mu\nu}^{\ \ \, \alpha}  
={1\over 2} ( 
 \delta_{\nu}^{\, \alpha}  \partial_{\mu} \phi 
 +  \delta_{\mu}^{\, \alpha} \partial_{\nu} \phi
 -S_{\mu\nu}S^{\alpha\beta}\partial_{\beta} \phi ),   \label{prima}\\
& & \hspace{-1cm}  
(2) \
 \Gamma_{\mu\nu}^{\ \ \, a} 
 =-{1\over 2}\tau^{-1} e^{\phi} \rho^{ab}S_{\mu\nu} \partial_{b} \phi, \label{facie}\\
& & \hspace{-1cm} 
(3) \ 
\Gamma_{\mu a}^{\ \ \, \nu} =\Gamma_{a \mu}^{\ \ \, \nu} 
= {1\over 2}\delta_{\mu}^{\, \nu} \partial_{a} \phi,    \label{remain}\\
& & \hspace{-1cm} 
(4) \ 
\Gamma_{0 a}^{\ \ \, b} =\Gamma_{a 0}^{\ \ \, b}
={1\over 2}\tau^{-1} \delta_{a}^{\, b} 
+ {1\over 2} \rho^{bc}  \partial_{\tau}  \rho_{ac},   \label{side2}\\
 & & \hspace{-1cm} 
(5) \ 
\Gamma_{R a}^{\ \ \, b}= \Gamma_{a R}^{\ \ \, b}
= {1\over 2} \rho^{bc}  \partial_{R}  \rho_{ac},       \label{defend}\\
 & & \hspace{-1cm} 
(6) \ 
\Gamma_{ab}^{\ \ \, 0}
= {1\over 2}  e^{-\phi} \rho_{ab}+ {1\over 2}\tau   e^{-\phi}  
(\partial_{\tau}  \rho_{ab} -  \partial_{R}  \rho_{ab} ),   \label{let}\\
 & & \hspace{-1cm} 
(7) \ 
\Gamma_{ab}^{\ \ \, R}=-{1\over 2}  e^{-\phi} \rho_{ab}
-{1\over 2}\tau   e^{-\phi}  \partial_{\tau}  \rho_{ab},    \label{coverage}\\
 & & \hspace{-1cm} 
(8) \ 
\Gamma_{ab}^{\ \ \, c}=:\Gamma_{ab}^{(2)c} 
= {1\over 2}\gamma^{cd} \big( 
\partial_{a}\gamma_{bd} +\partial_{b}\gamma_{ad} 
 - \partial_{d}\gamma_{ab}  \big)         \nonumber\\
& & \hspace{0.3cm}
={1\over 2}\rho^{cd} \big( 
\partial_{a}\rho_{bd} +\partial_{b}\rho_{ad}  - \partial_{d}\rho_{ab}  \big), \label{cutoff} \\
 & & \hspace{-1cm} 
(9) \ 
\Gamma_{ab}^{\ \ \, a}=:\Gamma_{ab}^{(2)a} =0. \label{cry}
\eea
Notice that in the second line of  Eq. (\ref{cutoff}), the coordinate condition
\be
{\partial \tau \over \partial Y^{a}}=0                             
\ee
was used. The Ricci tensors are given by
\bea  
& & \hspace{-1cm} 
(10) \ 
 {\rm R}_{\mu\nu} = \partial_{\mu} 
 ( \Gamma_{\alpha \nu}^{\ \ \, \alpha}  +  \Gamma_{a \nu}^{\ \ \, a} ) 
 -\partial_{\alpha} \Gamma_{\mu \nu}^{\ \ \, \alpha} 
- \partial_{a} \Gamma_{\mu \nu}^{\ \ \, a} \no\\
& &  \hspace{-0.7cm} 
+\Gamma_{\mu \beta}^{\ \ \, \alpha} \Gamma_{\alpha \nu}^{\ \ \, \beta}  
+\Gamma_{\mu a}^{\ \ \, \alpha} \Gamma_{\alpha \nu}^{\ \ \, a} 
+ \Gamma_{\mu \alpha}^{\ \ \, a} \Gamma_{a \nu}^{\ \ \, \alpha}
+\Gamma_{\mu a}^{\ \ \, b} \Gamma_{b \nu}^{\ \ \, a}\no\\
& & \hspace{-0.7cm} 
-  ( \Gamma_{\beta \alpha}^{\ \ \, \beta}  +  \Gamma_{a \alpha}^{\ \ \, a} ) 
 \Gamma_{\mu \nu}^{\ \ \, \alpha}  
 - ( \Gamma_{\beta a}^{\ \ \, \beta}  +  \Gamma_{b a}^{\ \ \, b} )
 \Gamma_{\mu \nu}^{\ \ \, a},                 \label{pen}
\eea
\bea 
& & \hspace{-1cm} 
(11) \ 
 {\rm R}_{0 a} = \partial_{0} 
 ( \Gamma_{\alpha a}^{\ \ \, \alpha}  +  \Gamma_{b a}^{\ \ \, b} ) 
 - \partial_{\alpha} \Gamma_{0 a}^{\ \ \, \alpha} 
 - \partial_{b} \Gamma_{0 a}^{\ \ \, b} \no\\
& &  \hspace{-0.7cm} 
+\Gamma_{0\beta}^{\ \ \, \alpha} \Gamma_{\alpha a}^{\ \ \, \beta}  
+\Gamma_{0 b}^{\ \ \, \alpha} \Gamma_{\alpha a}^{\ \ \, b} 
+\Gamma_{0\alpha}^{\ \ \, b} \Gamma_{b a}^{\ \ \, \alpha}  
+\Gamma_{0 c}^{\ \ \, b} \Gamma_{b a}^{\ \ \, c}  \no\\
& &  \hspace{-0.7cm} 
- ( \Gamma_{\beta \alpha}^{\ \ \, \beta}  +  \Gamma_{b \alpha}^{\ \ \, b} ) 
\Gamma_{0 a}^{\ \ \, \alpha}
-( \Gamma_{\beta b}^{\ \ \, \beta}  +  \Gamma_{c b}^{\ \ \, c} ) 
\Gamma_{0 a}^{\ \ \, b},                     \label{pencil}
\eea
\bea 
& & \hspace{-1cm} 
(12) \ 
 {\rm R}_{R a} = \partial_{R} 
 ( \Gamma_{\alpha a}^{\ \ \, \alpha}  +  \Gamma_{b a}^{\ \ \, b} ) 
 - \partial_{\alpha} \Gamma_{R a}^{\ \ \, \alpha} 
 - \partial_{b} \Gamma_{R a}^{\ \ \, b} \no\\
& &  \hspace{-0.7cm} 
+\Gamma_{R\beta}^{\ \ \, \alpha} \Gamma_{\alpha a}^{\ \ \, \beta}  
+\Gamma_{R b}^{\ \ \, \alpha} \Gamma_{\alpha a}^{\ \ \, b} 
+\Gamma_{R\alpha}^{\ \ \, b} \Gamma_{b a}^{\ \ \, \alpha}  
+\Gamma_{R c}^{\ \ \, b} \Gamma_{b a}^{\ \ \, c}  \no\\
& &  \hspace{-0.7cm} 
- ( \Gamma_{\beta \alpha}^{\ \ \, \beta}  +  \Gamma_{b \alpha}^{\ \ \, b} ) 
\Gamma_{R a}^{\ \ \, \alpha}
-( \Gamma_{\beta b}^{\ \ \, \beta}  +  \Gamma_{c b}^{\ \ \, c} ) 
\Gamma_{R a}^{\ \ \, b},                     \label{sharp2}
\eea
\bea  
& & \hspace{-1cm} 
(13) \ 
 {\rm R}_{ac} = \partial_{a} 
 ( \Gamma_{\mu c}^{\ \ \, \mu}  +  \Gamma_{b c}^{\ \ \, b} ) 
 - \partial_{\mu} \Gamma_{ac}^{\ \ \, \mu} 
 - \partial_{b} \Gamma_{ac}^{\ \ \, b} \no\\
& &  \hspace{-0.7cm} 
+\Gamma_{a\nu}^{\ \ \, \mu} \Gamma_{\mu c}^{\ \ \, \nu}  
+\Gamma_{ab}^{\ \ \, \mu} \Gamma_{\mu c}^{\ \ \, b} 
+\Gamma_{a\mu}^{\ \ \, b} \Gamma_{b c}^{\ \ \, \mu}  
+\Gamma_{a d}^{\ \ \, b} \Gamma_{b c}^{\ \ \, d}  \no\\
& &  \hspace{-0.7cm} 
- ( \Gamma_{\alpha \mu}^{\ \ \, \alpha}  +  \Gamma_{b \mu}^{\ \ \, b} ) 
\Gamma_{ac}^{\ \ \, \mu}
-( \Gamma_{\alpha b}^{\ \ \, \alpha}  +  \Gamma_{d b}^{\ \ \, d} ) 
\Gamma_{ac}^{\ \ \, b}.                     \label{ink}
\eea
The vacuum Einstein's equations are as follows:
\bea 
& & \hspace{-0.8cm}
1. \ {\rm R}_{00} = 0 \Rightarrow \no\\
& &  \hspace{-0.7cm}
\partial_{\tau}\phi - {\tau\over 4} \rho^{ab}\rho^{cd}
(\partial_{\tau}\rho_{ac})(\partial_{\tau}\rho_{bd}) 
+ {1\over 2}\tau^{-1} = 0,   \label{mundus}
\eea
\bea
&  & \hspace{-0.8cm}
2. \ {\rm R}_{0 R} = 0 \Rightarrow   \no\\
& & \hspace{-0.7cm}
- \partial_{\tau}^{2}\phi +2\partial_{\tau}\partial_{R}\phi 
-\tau^{-1} \partial_{\tau} \phi   
+  \tau^{-1} \partial_{a} 
( e^{\phi} \rho^{ab} \partial_{b}  \phi )            \no\\
& & \hspace{-0.4cm}
+ {1\over 2} \rho^{ab}\rho^{cd}
(\partial_{\tau}\rho_{ac})(\partial_{R}\rho_{bd}) =0, \label{round}
\eea
\bea
&  &  \hspace{-0.8cm}
3. \ {\rm R}_{R R} = 0 \Rightarrow   \no\\
& & \hspace{-0.7cm}
-\partial_{\tau}^{2}\phi + 2 \partial_{\tau}\partial_{R}\phi 
-\tau^{-1} \partial_{\tau} \phi   
+\tau^{-1} \partial_{R} \phi \no\\
& & \hspace{-0.7cm}
+ \tau^{-1} \partial_{a} 
( e^{\phi} \rho^{ab} \partial_{b}  \phi )   
+ {1\over 2} \rho^{ab}\rho^{cd}
(\partial_{R}\rho_{ac})(\partial_{R}\rho_{bd})    \no\\
& &  \hspace{-0.4cm}
=0,                \label{erasmus}
\eea
\bea
&  & \hspace{-0.8cm}
4. \ {\rm R}_{0 a} = 0 \Rightarrow   \no\\
& & \hspace{-0.4cm}
 \partial_{\tau}\partial_{a}\phi  - \tau^{-1}\partial_{a}\phi
 -\partial_{b}( \rho^{bc}\partial_{\tau}\rho_{ac})    \no\\
& & \hspace{-0.4cm}
 +{1\over 2}\rho^{bc}\rho^{de}(\partial_{\tau}\rho_{bd})(\partial_{a}\rho_{ce})
 =0,                  \label{donqui}
\eea
\bea
&  &  \hspace{-0.8cm}
5. \ {\rm R}_{R a} = 0 \Rightarrow   \no\\
& & \hspace{-0.4cm}
 \partial_{R}\partial_{a}\phi 
 -\partial_{b}( \rho^{bc}\partial_{R}\rho_{ac})\no\\
& & \hspace{-0.4cm}
 +{1\over 2}\rho^{bc}\rho^{de}(\partial_{R}\rho_{bd})(\partial_{a}\rho_{ce})=0,    \label{donde}
\eea
\bea
&  & \hspace{-0.8cm}
6. \ {\rm R}_{ac} = 0 \Rightarrow   \no\\
& & \hspace{-0.4cm}
-  {\rm R}_{(2)ac} + {1\over 2} (\partial_{a} \phi )(\partial_{c} \phi )
+ \partial_{a}\partial_{c}  \phi -\Gamma_{ac}^{(2)b } \partial_{b} \phi  \no\\
& & \hspace{-0.4cm}
-{1\over 2} e^{-\phi}  ( \partial_{\tau}\rho_{ac}-\partial_{R}\rho_{ac} )
+\tau  e^{-\phi} \partial_{R}\partial_{\tau}  \rho_{ac}  \no\\
& & \hspace{-0.4cm}
-{\tau\over 2} e^{-\phi}   \partial_{\tau}^{2}\rho_{ac}
+{\tau\over 4} e^{-\phi}  \rho^{bd} (\partial_{\tau}\rho_{cd} )
 ( \partial_{\tau}\rho_{ab}-\partial_{R}\rho_{ab} ) \no\\
& & \hspace{-0.4cm}
+{\tau\over 4} e^{-\phi}  \rho^{bd} (\partial_{\tau}\rho_{ad} )
 ( \partial_{\tau}\rho_{bc}-\partial_{R}\rho_{bc} )   \no\\
& & \hspace{-0.4cm}
-{\tau\over 4} e^{-\phi}  \rho^{bd}  \big\{ 
(\partial_{\tau}\rho_{ab})(\partial_{R}\rho_{cd} )
+(\partial_{\tau}\rho_{bc})(\partial_{R}\rho_{ad})  \big\} \no\\
& & \hspace{-0.4cm}
=0, \label{beach}
\eea
where $\Gamma_{ac}^{(2)b }$ is the Levi-Civita connection given by 
Eq. (\ref{cutoff}), and ${\rm R}_{(2)ac}$ is the Ricci tensor of $N_{2}$.

\subsection{Equivalence of Einstein's Equations and Hamilton's Equations of Motion
in Privileged Coordinates}{\label{a:zwei}}

Now, Eqs. (\ref{mundus}), $\cdots$, (\ref{beach}) will be shown to be equivalent to
Hamilton's equations of motion summarized in Section \ref{a:main}.

\noindent
(a) \ Notice that from Eq. (\ref{veryeasy}),
\be \hspace{-0.5cm} 
\partial_{\tau}\rho_{ab}=2 \tau^{-1}\rho_{a c}\rho_{b d}\pi^{cd}
+\partial_{R}\rho_{ab},       \label{night}
\ee
one finds
\be
{\tau\over 4}\rho^{ab}\rho^{cd}
(\partial_{\tau}\rho_{ac})(\partial_{\tau}\rho_{bd})  - {1\over 2}\tau^{-1}
={\mathcal H} - \tau^{-1}.                        \label{set}
\ee
where ${\mathcal H}$ is given by Eq. (\ref{trueham}). Therefore, Eq. (\ref{mundus}) becomes
\be
\partial_{\tau}\phi = {\mathcal H} - \tau^{-1}.   \label{steinbeck}
\ee
However, one has
\be
\partial_{\tau}\phi =\partial_{\tau} \ln (-h), \label{energy}
\ee
by taking the $\tau$ derivative of both sides of Eq. (\ref{respect}), 
so Eq. (\ref{steinbeck}) becomes
\be
\partial_{\tau}  {\ln}(-h) ={\mathcal H} -\tau^{-1},    \label{rado}
\ee
which is identical to Eq. (\ref{hstar}) in Section \ref{a:main}. 

\noindent 
(b) \ 
If Eq. (\ref{erasmus}) is subtracted from Eq. (\ref{round}), and if the result is
multiplied by $\tau$, one finds
\be
-\partial_{R} \phi + {\tau\over 2} \rho^{ab}\rho^{cd}
(\partial_{\tau}\rho_{ac} 
-\partial_{R}\rho_{ac}) (\partial_{R}\rho_{bd}) =0.                \label{dark} 
\ee
If the defining equation of $\pi_{R}$ in Eq. (\ref{pira}) is used, then 
the $R$ derivative of  Eq.  (\ref{respect}) becomes
\be
\partial_{R} \phi = - \pi_{R}.                \label{bira}
\ee
By  Eqs.  (\ref{veryeasy}) and (\ref{bira}),  Eq. (\ref{dark}) 
becomes
\be
\pi_{R} +\pi^{ab}\partial_{R}\rho_{ab}  =0,        \label{grandure}
\ee
which is identical to  Eq. (\ref{radi}) in Section \ref{a:main}.

\noindent
(c) \ 
If Eq. (\ref{donde}) is subtracted from Eq. (\ref{donqui}), and then multiplied 
by $\tau$, one finds
\bea
& &  \hspace{-1cm} 
\tau\partial_{a} \partial_{\tau}\phi  -\tau\partial_{a} \partial_{R}\phi 
 - \partial_{a}\phi
 -\tau\partial_{b} \{ \rho^{bc} (\partial_{\tau}\rho_{ac} -\partial_{R}\rho_{ac}) \} \no\\
& & \hspace{-1cm} 
+{\tau\over 2}\rho^{bc}\rho^{de} (\partial_{a}\rho_{ce}) (\partial_{\tau}\rho_{bd} -\partial_{R}\rho_{bd}) =0.                              \label{pale}
\eea
If Eq. (\ref{piwa}) is used, the $Y^{a}$ derivative of Eq. (\ref{respect}) becomes
\be
\partial_{a}\phi = - \tau^{-1}\pi_{a}.               \label{heine}
\ee
If the $Y^{a}$ derivatives of  Eqs.  (\ref{steinbeck}) and (\ref{bira})
are used, together with the coordinate condition
\be
\partial_{a}\tau=0,
\ee
then the first three terms of  Eq. (\ref{pale}) become
\be
\tau \partial_{a}\partial_{\tau} \phi  -\tau \partial_{a}\partial_{R}\phi 
 - \partial_{a}\phi            
 ={\partial_{a}}\{ \tau ({\mathcal H} + \pi_{R}) \}+ \tau^{-1}\pi_{a}.   \label{sky}
\ee
The fourth and the fifth terms of  Eq. (\ref{pale}) can be written as
\bea
& &   \hspace{-1cm} 
({\rm i}) 
\ -\tau\partial_{b} \{ \rho^{bc} (\partial_{\tau}\rho_{ac} -\partial_{R}\rho_{ac}) \} 
=-2 \partial_{b}  (\rho_{ac}\pi^{bc}),  \label{wash}\\
& &   \hspace{-1cm} 
({\rm ii})\ 
{\tau\over 2}\rho^{bc}\rho^{de} (\partial_{a}\rho_{ce}) (\partial_{\tau}\rho_{bd} -\partial_{R}\rho_{bd})
=\pi^{bc}\partial_{a} \rho_{bc},       \label{light}
\eea
respectively. By  Eqs. (\ref{sky}), (\ref{wash}), and (\ref{light}),  Eq.  (\ref{pale}) becomes 
\be   
\tau^{-1}{\pi}_{a}
= - \pi^{bc}{\partial_{a}} \rho_{bc}  
+2{\partial_{b}} (\pi^{bc}\rho_{ac})   
- {\partial_{a}} \{ \tau ( {\mathcal H} + \pi_{R}) \},           \label{lower}
\ee
which is identical to  Eq.  (\ref{angko}) in Section \ref{a:main}. 

\noindent 
(d) \ 
If the trace of  Eq.  (\ref{beach}) with $\rho^{ac}$ is taken, then many terms cancel out and the equation becomes
\be
\tau {\rm R}_{(2)} - {1 \over 2} \rho^{ab}(\partial_{a} \phi)(\partial_{b} \phi)
-\partial_{a}  ( \rho^{ab} \partial_{b} \phi) =0,     \label{east}
\ee
where ${\rm R}_{(2)}$ is the scalar curvature of $N_{2}$. 
Equation (\ref{east}) can be written as, if Eq. (\ref{heine}) is used,
\be
\tau {\rm R}_{(2)} - {1\over 2}\tau^{-2}\rho^{ab}\pi_{a}
\pi_{b} + { \partial_{a}} (\tau^{-1}\rho^{ab}\pi_{b}) = 0,  \label{book}
\ee
which is identical to  Eq. (\ref{topo})  in Section \ref{a:main}. 

\noindent 
(e) \
Now, examine  Eq.  (\ref{beach}).  Notice that
the 2-dimensional covariant derivative of $ \partial_{c}  \phi $ is given by
\be
\nabla_{a}^{(2)} \partial_{c}  \phi = \partial_{a}\partial_{c}\phi 
-\Gamma_{ac}^{(2)b } \partial_{b}\phi.        \label{orga}
\ee
If the $\tau$ and $R$ derivatives of Eq. (\ref{veryeasy}) are taken, 
then one finds
\bea
& &   \hspace{-1cm} 
({\rm i}) \ 
\partial_{\tau}^{2} \rho_{ac}   \nonumber\\
& &  \hspace{-1cm} 
=2 \tau^{-1}  \rho_{ab} \rho_{cd} (\partial_{\tau}\pi^{bd})
+ 2  \tau^{-1}  (\partial_{R}\rho_{ab}) \rho_{cd}\pi^{bd}   \no\\
& &  \hspace{-0.8cm} 
+ 2  \tau^{-1}  (\partial_{R}\rho_{cd}) \rho_{ab}\pi^{bd} 
+2 \tau^{-1}  \rho_{ab} \rho_{cd}  (\partial_{R} \pi^{bd})  \no\\
& &  \hspace{-0.8cm} 
+\partial_{R}^{2}\rho_{ac} 
-2 \tau^{-2}  \rho_{ab} \rho_{cd}\pi^{bd} 
+2 \tau^{-1}  (\partial_{\tau}\rho_{ab}) \rho_{cd} \pi^{bd}   \no\\
& &  \hspace{-0.8cm} 
+2 \tau^{-1}  (\partial_{\tau}\rho_{cd}) \rho_{ab} \pi^{bd},   \label{likely}\\
& &   \hspace{-1cm} 
({\rm ii}) \
\partial_{R}\partial_{\tau} \rho_{ac}\nonumber\\
& &  \hspace{-1cm} 
=2  \tau^{-1} (\partial_{R}\rho_{ab})\rho_{cd}\pi^{bd} 
+\partial_{R}^{2} \rho_{ac}     \no\\
& &  \hspace{-0.8cm} 
+ 2  \tau^{-1} \rho_{ab}  (\partial_{R}\rho_{cd}) \pi^{bd}  
+2 \tau^{-1}\rho_{ab} \rho_{cd}  (\partial_{R} \pi^{bd} ).   \label{perhaps}
\eea
If Eqs. (\ref{veryeasy}), (\ref{orga}), (\ref{likely}), and (\ref{perhaps})
are plugged into Eq. (\ref{beach}), then the equation becomes
\bea
& & \hspace{-1.2cm}
-  {\rm R}_{(2)ac} + {1\over 2} (\partial_{a} \phi )(\partial_{c} \phi )
+ \nabla_{a}^{(2)}\partial_{c}  \phi  \no\\
& & \hspace{-1.2cm}
-e^{-\phi}   \rho_{ab}   \rho_{cd} \partial_{\tau}\pi^{bd}
+ e^{-\phi}   \rho_{ab}   \rho_{cd} \partial_{R}\pi^{bd}
+{\tau\over 2} e^{-\phi}\partial_{R}^{2}\rho_{ac}  \no\\
& & \hspace{-1.2cm}
-\tau^{-1} e^{-\phi}  \rho_{ad}   \rho_{ci}   \rho_{bj}  \pi^{ij}\pi^{bd}
-\tau^{-1} e^{-\phi}  \rho_{cd}   \rho_{ai}   \rho_{bj}  \pi^{ij}\pi^{bd} \no\\
& & \hspace{-1.2cm}
-{\tau\over 2} e^{-\phi} \rho^{bd} (\partial_{R}\rho_{ab})(\partial_{R}\rho_{cd})
=0.                  \label{girl}
\eea
By contracting  Eq. (\ref{girl}) with $e^{\phi}\rho^{ak}\rho^{cl}$ and using  Eq. (\ref{heine}), after some algebra,  Eq. (\ref{girl})  becomes
\bea
& & \hspace{-0.6cm}
\partial_{\tau}\pi^{ab}
=-2\tau^{-1}\rho_{c d}\pi^{a c}\pi^{b d}  + \partial_{R}\pi^{ab}   \no\\
& &  \hspace{-0.7cm}
-{\tau\over 2}\rho^{a i}\rho^{b j}\rho^{c k} 
(\partial_{R}\rho_{i c})(\partial_{R}\rho_{j k})  
+{\tau\over 2}\rho^{ac}\rho^{b  d} (\partial_{R}^{2}\rho_{c d}) \no\\
& &  \hspace{-0.7cm}
+{1\over 2}e^{\phi} \tau^{-2}\rho^{ac}\rho^{b  d} \pi_{c}\pi_{d}  
-e^{\phi}\rho^{ac}\rho^{b  d} \nabla_{c}^{(2)}  (\tau^{-1} \pi_{d}) \no\\
& &  \hspace{-0.7cm}
-e^{\phi} \rho^{a c}\rho^{b d} {\rm R}_{(2)c d}.          \label{stainless}
\eea
The last three terms in  Eq. (\ref{stainless}) 
become
\bea
& &  \hspace{-1cm}
{1\over 2}e^{\phi} \tau^{-2}\rho^{ac}\rho^{b  d} \pi_{c}\pi_{d}  
-e^{\phi}\rho^{ac}\rho^{b  d} \nabla_{c}^{(2)}  (\tau^{-1} \pi_{d}) \no\\
& &  \hspace{-1cm}
-e^{\phi} \rho^{a c}\rho^{b d} {\rm R}_{(2)c d},   \no\\
& & \hspace{-1cm}
=2h \rho^{a c}\rho^{b d} \{ 
{\rm R}_{(2)c d} - {1\over 2}\tau^{-2}\pi_{c}\pi_{d} 
+\nabla_{c}^{(2)}  (\tau^{-1} \pi_{d}) \},          \label{silver}
\eea
where $e^{\phi}$ was replaced with $-2h$. If Eq. 
(\ref{silver}) is substituted into Eq.  (\ref{stainless}), then
Eq. (\ref{stainless}) becomes identical to 
Eq. (\ref{getzen}) in Section \ref{a:main}.\\

\noindent
(f)\ 
Now, examine  Eq.  (\ref{donqui}). If Eq. (\ref{heine}) is used,  
then the first two terms of  Eq. (\ref{donqui}) become
\be
\partial_{\tau}\partial_{a}\phi -\tau^{-1}\partial_{a} \phi  
=-\tau^{-1}\partial_{\tau}\pi_{a}+2\tau^{-2}\pi_{a}.    \label{still}
\ee
By  Eq. (\ref{veryeasy}), the third and the fourth terms of  Eq. (\ref{donqui}) become
\bea 
& &  \hspace{-1.3cm}
({\rm i}) \, -\partial_{b}( \rho^{bc}\partial_{\tau}\rho_{ac})  \!
=\! -\tau^{-1} \partial_{b} (
2 \pi^{bc}\rho_{ac} +\tau \rho^{bc}\partial_{R}\rho_{ac} ),  \label{canada}\\
& & \hspace{-1.3cm}
({\rm ii})\, 
{1\over 2}\rho^{bc}\rho^{de}(\partial_{\tau}\rho_{bd})(\partial_{a}\rho_{ce})\no\\
& & \hspace{0.2cm}
=\tau^{-1}(  \pi^{ce} + {\tau \over 2}\rho^{bc}\rho^{de}\partial_{R}\rho_{bd})
(\partial_{a}\rho_{ce}),             \label{vancouver}
\eea 
respectively. Thus,  Eq.  (\ref{donqui}) becomes, after multiplying
it by $\tau$, 
\bea
& & \hspace{-0.5cm}
-\partial_{\tau}\pi_{a}+2\tau^{-1}\pi_{a} 
-\partial_{b} (
2 \pi^{bc}\rho_{ac} +\tau\rho^{bc}\partial_{R}\rho_{ac} )    \nonumber\\
& & \hspace{-0.5cm}
+(  \pi^{ce} + {\tau\over 2}\rho^{bc}\rho^{de}\partial_{R}\rho_{bd})
(\partial_{a}\rho_{ce}) =0,               \label{pearl}
\eea
which is identical to  Eq.  (\ref{parpi}) in Section \ref{a:main}.\\

\noindent
(g)\ 
Now, examine  Eq.  (\ref{round}).  If Eq. (\ref{topo}) is used, then  Eq. (\ref{pitau}) becomes
\bea
& & 
\pi_{\tau} = -{\mathcal H} +2\partial_{R} \ln (-h)         \nonumber\\
& & \hspace{0.5cm}
=-\partial_{\tau} \phi -\tau^{-1} +2 \partial_{R} \phi,   \label{buck}
\eea
where Eq.  (\ref{steinbeck}) is used in the second line. Thus, 
\be \hspace{-0.4cm}
-\partial_{\tau} \phi + 2 \partial_{R}\phi =  \pi_{\tau} + \tau^{-1}.  \label{find}
\ee
If the $\tau$ derivative of both sides of  Eq. (\ref{find}) is taken, then
\be \hspace{-0.4cm}
- \partial_{\tau}^{2}\phi +2 \partial_{\tau}\partial_{R}\phi 
=\partial_{\tau} \pi_{\tau} - \tau^{-2},    \label{cent}
\ee
which is the first two terms of  Eq. (\ref{round}).
The third term of  Eq. (\ref{round}) becomes
\bea
& &  \hspace{-1cm}
-\tau^{-1} \partial_{\tau} \phi 
=- \tau^{-1} {\mathcal H} + \tau^{-2}   \no\\
& & \hspace{-0.4cm}
={1\over 2}\tau^{-2} 
- \tau^{-2} \rho_{ab}\rho_{cd} \pi^{ac} \pi^{bd} 
-\tau^{-1} \pi^{ac} \partial_{R} \rho_{ac}  \no\\
& & \hspace{-0.4cm}
-{1\over 4}\rho^{ab}\rho^{cd} (\partial_{R}\rho_{ac})(\partial_{R}\rho_{bd}), \label{edition}
\eea
where $ {\mathcal H} $ given by  Eq. (\ref{trueham}) was used. 
The fourth and the fifth terms of  Eq. (\ref{round}) can be written as
\bea
& &  \hspace{-1.2cm}
({\rm i}) \ \tau^{-1} \partial_{a} 
( e^{\phi} \rho^{ab} \partial_{b}  \phi )  
=2 \tau^{-2} \partial_{a} (h \rho^{ab} \pi_{b}),      \label{deep}\\
& & \hspace{-1.2cm}
({\rm ii}) \ {1\over 2} \rho^{ab}\rho^{cd}
(\partial_{\tau}\rho_{ac})(\partial_{R}\rho_{bd})   \no\\
& &  \hspace{-0.6cm}
=\tau^{-1} \pi^{bd} \partial_{R}\rho_{bd} 
+  {1\over 2} \rho^{ab}\rho^{cd}
(\partial_{R}\rho_{ac}) (\partial_{R}\rho_{bd}).       \label{human}
\eea
If Eqs.  (\ref{cent}), (\ref{edition}), (\ref{deep}), and 
(\ref{human}) are plugged into  Eq. (\ref{round}), then 
\bea
& &  \hspace{-1cm} 
\partial_{\tau} \pi_{\tau} - {1\over 2} \tau^{-2}
+{1\over 4} \rho^{ab}\rho^{cd}
(\partial_{R}\rho_{ac}) (\partial_{R}\rho_{bd})   \no\\
& & \hspace{-0.7cm} 
- \tau^{-2} \rho_{ab}\rho_{cd} \pi^{ac} \pi^{bd} 
+2 \tau^{-2} \partial_{a} (h \rho^{ab} \pi_{b})=0, \label{mexican}
\eea
which is just  Eq.  (\ref{volvopitau}) in Section \ref{a:main}.\\

\noindent
(h)\ 
Finally, examine  Eq. (\ref{donde}). If Eq. (\ref{bira}) is used, then
\be \hspace{0cm} 
\partial_{a}\pi_{R}
 -{1\over 2}\rho^{bc}\rho^{de}(\partial_{R}\rho_{bd})(\partial_{a}\rho_{ce})
 +\partial_{b}( \rho^{bc}\partial_{R}\rho_{ac})=0. \!\!\!   \label{patata}
\ee
This equation can be reproduced 
from the evolution equation of $\pi_{a}$ given by  Eq. (\ref{parpi}),
the defining equation of the physical angular momentum given by (\ref{angko}), 
and the integrability condition given by Eq. (\ref{partautau}).
Notice that  Eq. (\ref{partautau}) can be written as
\be
\partial_{\tau}\pi_{a} - \tau^{-1} \pi_{a} 
+ \tau \partial_{a} {\mathcal H} = 0. \label{insalata}
\ee
If the right hand side of  Eqs. (\ref{parpi}) is substituted for $\partial_{\tau}\pi_{a}$ in the above equation, then 
\bea
& & 
\tau^{-1} \pi_{a}  
+( \pi^{ce} + {\tau\over 2}\rho^{bc}\rho^{de}\partial_{R}\rho_{bd}) 
(\partial_{a}\rho_{ce})            \nonumber\\
& &
-\partial_{b} (
2 \pi^{bc}\rho_{ac} +\tau\rho^{bc}\partial_{R}\rho_{ac} ) 
+  \tau \partial_{a} {\mathcal H} = 0.          \label{shell}
\eea
Now, if the right hand side of  Eq. (\ref{angko}) is substituted 
for $\tau^{-1} \pi_{a}$ in Eq. (\ref{shell}), then Eq.  (\ref{shell}) becomes identical to  Eq. (\ref{patata}). 

Therefore, Einstein's equations ${\rm R}_{AB} =0$ of the spacetime
metric given by Eq. (\ref{yoonspacetime}) 
in privileged coordinates are identical to the set of first-order 
equations that are obtained from the Hamiltonian reduction and 
summarized in Section \ref{a:main}. This proves that the Hamiltonian reduction discussed in this paper
is a consistent procedure.

\section{Discussion}{\label{a:discussion}}
\setcounter{equation}{0}

The most important aspect of the Hamiltonian reduction is
the introduction of the privileged coordinates $(\tau, R, Y^{a})$.
It is the very existence of these privileged coordinates 
that not only allows the Hamiltonian reduction, 
but also enables informations such as 
physical Hamiltonian and momentum of the gravitational fields 
to be extracted from the constraints by solving them explicitly.
Therefore, it is appropriate to review the meaning of privileged 
coordinates and examine their limitations.
Basically, privileged coordinates are local in nature,
although they  may be extended to the sufficiently large patch of 
spacetimes under consideration. 
The area element $\tau$ of the spatial 
cross-section of the out-going null hypersurface is chosen as the physical time.
On a $\tau={\rm constant}$  hypersurface $\Sigma_{\tau}$, the equipotential surface
defined by the potential function $R$ in  Eq. (\ref{iden}) defines 
the surface $N_{2}$ of constant radius,  on which the coordinates $Y^{a}$ 
are chosen such that the families of $Y^{a}={\rm constant}$ subsurfaces are
orthogonal to the $R={\rm constant}$ equipotential surface, i.e.,
$A_{+}^{\ a}=0$.
Any legitimate notion of time requires that the time function 
be an invariant scalar on a 3-dimensional hypersurface, and the fact 
that $\tau$ is such a scalar function with weight $0$ 
is clear from the fact that the ``shift" $A_{+}^{\ a}$  is 
zero in privileged coordinates. 

By the same reason, the function $R$ in  Eq. (\ref{iden}) becomes
a scalar function too and therefore, it was chosen as the legitimate 
physical radial coordinate.  
Thus, $(\tau, R, Y^{a})$ are the privileged coordinates relevant 
for the description of the 4-dimensional spacetime
in which the metric is written as
\begin{equation}
d s^{2}=-2h d s^{2}_{(2), {\rm flat}} + \tau\rho_{ab} dY^{a}  dY^{b},  \label{kino}
\end{equation}
where $d s^{2}_{(2),{\rm flat}}$ is the flat (1+1)-dimensional metric
\begin{equation}
d s^{2}_{(2), {\rm flat}} =2dRd\tau + dR^{2},   \label{awaken}
\end{equation}
and the conformal factor is $-2h>0$ such that $\tau={\rm constant}$ is 
spacelike. It may be appropriate to mention that privileged coordinates do not
cover the whole spacetime in general.
For example, the area element $\tau$ of the null hypersurface might start to decrease,
as is the case of the recollapsing null hypersurface surrounding strong attractive
sources.  Then, at some moment, 
\begin{equation}
d \tau =0,          \label{serious}
\end{equation}
violating the monotonicity of $\tau$, and $\tau$
becomes a bad coordinate at that moment. However, a coordinate singularity
of this kind is a familiar one, as it also appears in mini-superspace cosmologies
where the volume of a 3-dimensional hypersurface, 
typically chosen as the physical time,
suffers the same difficulty at the moment of maximum expansion.

In addition to the monotonicity of $\tau$, the condition that {\it no caustics} exist 
on the null hypersurface is also necessary for the validity of privileged coordinates.
For instance, at caustics such as the vertex of a lightcone where the area element
$\tau$ becomes
\be
\tau=0, 
\ee
the physical Hamiltonian density in Eq. (\ref{trueham}) diverges as
\be
H \sim  {1\over \tau}\rightarrow \infty.        \label{dive}
\ee
The singularity of the physical Hamiltonian density should be regarded as 
a signal that privileged coordinates become bad coordinates at points such as 
caustics.

The condition of vanishing ``shift" is also a local condition.
If this condition is to hold globally, then
3-dimensional spacelike hypersurface $\Sigma_{\tau}$ must be factorizable such as
$N_{2}\times \mathbb{R}^{1}$ or  $N_{2}\times S^{1}$. 
If $\Sigma_{\tau}$ is not so, for example, $\Sigma_{\tau}=S^{3}$, then the ``shift" does not vanish in general. 
In this case, different privileged coordinates must be sought so as to make the Hamiltonian reduction possible. This work is in progress.
However, the fact that the spacetime metric in Eq. (\ref{kino}) is a general line element, {\it not} a particular
solution to Einstein's equations, should be stressed; therefore, it is not surprising that
the privileged coordinates do not cover the entire spacetime,
but  are limited to the region where $d\tau \neq 0$ and 
the ``shift" is vanishing. 

I would like to emphasize that
{\it all} the constraint equations $C_{\pm}=0$ and $C_{a}=0$
in privileged coordinates are solved in that the constraints simply
define the physical Hamiltonian and momentum densities in terms of 
only the physical degrees of freedom $(\rho_{ab}, \pi^{ab})$, 
as is summarized in  Eqs. (\ref{pitau}), (\ref{radi}), and (\ref{angko}). 
Although the present work deals with the case 
where the 4-dimensional spacetime is foliated into families of null and spacelike hypersurfaces whose intersection defines a compact two surface, it can be extended to include different foliations such as the null and timelike splittings and double null splitting\cite{dinverno80,smallwood83}. The case of null and timelike splittings can be studied by choosing the sign $-2h < 0$, but the price one has to pay is that 
the Hamiltonian and other canonical quantities are defined on a {\it timelike} hypersurface, rather than on a spacelike hypersurface. Double null case is a bit
delicate, but in principle, it can be studied by taking 
the limits $-2h \rightarrow 0^{\pm}$ of the null and spacelike splitting or null and timelike splitting.


\begin{acknowledgments}

This work is supported in part by Konkuk University (2013-A019-0133).

\end{acknowledgments}

\end{document}